\documentclass[nofootinbib,aps,pra,showpacs,groupedaddress]{revtex4}
\usepackage{mathtools}
\usepackage{amssymb}
\usepackage{graphics}
\usepackage[dvips]{graphicx}
\usepackage{graphicx}
\usepackage{color}
\usepackage{wrapfig}
\usepackage{subfigure}

\newcommand \beq{\begin{eqnarray}}
\newcommand \eeq{\end{eqnarray}}
\def\simge{\mathrel{%
       \rlap{\raise 0.511ex \hbox{$>$}}{\lower 0.511ex \hbox{$\sim$}}}}
\def\simle{\mathrel{
       \rlap{\raise 0.511ex \hbox{$<$}}{\lower 0.511ex \hbox{$\sim$}}}}

\begin{document}
\title{Ultrarelativistic heavy ion collisions: the first billion seconds}
\author{Gordon Baym}
\affiliation{\mbox{Department of Physics, University of Illinois, 1110
  W. Green Street, Urbana, IL 61801} \\
}

\date{\today}

\begin{abstract}
   I first review the early history of the ultrarelativistic heavy ion program, starting with the 1974 Bear Mountain Workshop, and the 1983 Aurora meeting of the U.S. Nuclear Science Committee, just one billion seconds ago, which laid out the initial science goals of an ultrarelativistic collider.   The primary goal, to discover the properties of nuclear matter at the highest energy densities, included finding new states of matter -- the quark-gluon plasma primarily -- and to use collisions to open a new window on related problems of matter in cosmology, neutron stars, supernovae, and elsewhere.   To bring out how the study of heavy ions and hot, dense matter in QCD has been fulfilling these goals.  I concentrate on a few topics, the phase diagram of matter in QCD, and connections of heavy ion physics to cold atoms, cosmology, and neutron stars.  
  
\end{abstract}

\maketitle

\section{Introduction}

\label{sec:1}

    Serious planning for the first ultrarelativistic heavy ion collider, RHIC, began in the summer of 1983, just about
$10^9$ seconds before the QM2015 conference, whence the title of this talk.   I would like first to review the early history of the ultrarelativistic heavy ion program,
to ask what we had in mind when planning and building RHIC, and incorporating heavy ions collisions into the LHC;  what were the scientific motivations; and to what extent have we shed light on the scientific issues and succeeded in fulfilling the promises of the program.   Much of this history was documented in the talk I prepared for the Quark Matter conference at Brookhaven in 2001 \cite{dreams}, so I will briefly recap the significant developments.   Since a non-trivial part of the justification for the program was its promise of making connections with other fields of physics, I would like also touch on connections that have and are still being made.
  
   The early 1974 Workshop at Bear Mountain, just north of New York City, on {\em BeV/nucleon collisions of
heavy ions,} was a pivotal event in the conception of heavy ion physics \cite{bearmountain}, since after this workshop physicists began to take seriously the possibility of using heavy ion collisions as a tool to study the properties of matter under extreme conditions of high energy and baryon densities, to ask whether was there a ``nuclear world quite different from the one we have learned to accept as familiar and stable?"  After some three decades of experiment which have successfully seen and continue to explore this {\em terra incognita}. we can answer this question very positively.  The question asked by T. D. Lee at the meeting -- could one see the restoration of broken symmetries, and create abnormal states of nuclear matter at high density in collisions -- foreshadowed restoration of broken chiral symmetry at high densities, albeit in the context of nucleons rather than quarks.  

  The Bear Mountain Workshop took place at the end of the period in which understanding of the basic structure of matter in terms of quarks was put on a firm footing. especially with the development of asymptotic freedom in 1973 \cite{qcd}.  Although quark matter was proposed as early as 1970 by Itoh in neutron stars and by Carruthers in 1973 \cite{earlyqm}, nuclear physicists at the time were more comfortable with hadronic based pictures of high densities, e.g.,  the Hagedorn resonance gas \cite{hagedorn} and the Walecka model \cite{walecka}.  The first mention of a deconfinement transition in dense matter was by Collins and Perry \cite{collins}, who suggested that matter is a ``quark soup" at high densities,  and then by Cabbibo and Parisi \cite{cabbibo}.  

   The 1970's saw considerable interest in developing facilities for heavy ion experiment among U.S., European, and Japanese physicists \cite{shoji}.  The Bevalac, which grew out of the Berkeley Bevatron, dates back to the early 1970's and was a principle training ground for subsequent European 
heavy ion research.  The AGS fixed target program at Brookhaven (BNL) was conceived at the time of Bear Mountain, and initiated in Jan. 1983, first with $^{16}$O beams, and then with Si and Au; experiments started in Oct. 1986 and dwindled down in the early 2000's after the start of RHIC physics.   In addition, in the late 1970's  the Institute for Nuclear Studies in Tokyo began developing the Numatron to accelerate ions as heavy as uranium up to energies of 1.3 GeV/A; unfortunately the machine was never approved for construction. 

  The first glimmer of RHIC traces back to the open meeting of the U.S. Nuclear Science Advisory Committee (NSAC)) held at Wells College in Aurora, N.Y. in July 1983, a meeting to help decide on the next major facility for nuclear physics in the U.S.     
  On the first day of this meeting the U.S. high energy community, meeting independently in Washington, decided to abandon the Colliding Beam Accelerator (or Isabelle, named after accelerator physicist J. Blewett's sailboat) being built at Brookhaven (BNL) in favor of the 200 GeV per nucleon  Desertron, or SSC, itself cancelled in 1993.   This decision presented our NSAC subcommittee on extreme states of nuclear matter \cite{committee} with an irresistible opportunity to build a relativistic heavy ion collider in the forlorn CBA tunnel, an idea J.D. Bjorken had first proposed informally at Fermilab in March 1983.   The committee's arguments, which I presented, met with wide acceptance in Aurora.  A proposal on the table for a Variable Energy Nuclear Synchrotron, or VENUS, fixed target machine at LBL, with energy up to 20 GeV/A, was abandoned.  Thus did RHIC enter the conceptual stage.    
  
   As we argued, the main goal for such a collider would be to discover the properties of extended nuclear matter at the highest densities.  What are the  gross features of its phase diagram, its equation of state and its entropy?    What are its dynamical properties, its excitations, and collective modes? 
How does it transport conserved quantities -- energy-momentum, baryons, etc.?  How does it stop hadronic and quark projectiles, and how does it dissipate energy?   How does it emit particles?  A ``frontier opportunity" would be ``discovering new states of matter, including a quark-gluon plasma."  Making a quark-gluon plasma was a goal, but not {\em the} goal.   The 1983 NSAC Long Range Plan \cite{nsaclrp} later summarized the scientific questions as, 
``What is the nature of nuclear matter at energy densities comparable
to those of the early universe?"  ``What are the new phenomena and physics associated with the
simultaneous collision of hundreds of nucleons at relativistic
energies?"  It stressed that the most outstanding opportunity opened by an ultrarelativistic heavy
ion collider is ``the creation of extended regions of nuclear matter at
energy densities beyond those ever created in the laboratory over
volumes far exceeding those excited in elementary particle
experiments and surpassed only in the early universe.''   
   
    Promised connections to other fields included, as we noted in Aurora, using heavy ion collisions to learn about QCD -- which at the time was still under the wing of high energy physics  -- to  see the deconfinement transition and determine its order, to see chiral symmetry restoration, and as well to learn the behavior of QCD at large distances.    Collisions, it was  hoped, might tell us about matter in the deep interiors of neutron stars, the nature of nuclear matter in supernovae; 
the role of the QCD confinement transition in cosmology -- e.g., possible production of 
black holes of order 0.01 solar masses in a first order phase transition -- and give insights into cosmic ray physics.   One would have a new arena to study many-body effects familiar in condensed matter, e.g, quasiparticles, broken symmetry states and their restoration.   

   Immediately after the Aurora meeting, BNL assembled a task force to set the design parameters of the new machine.  The choice of maximum beam energy, 100 GeV/A, within the constraints of the tunnel design, was driven by Bjorken's suggestion of producing jets that would propagate
through and thus probe the collision volume.   The importance of being able to vary the beam energy as well as the projectiles, e.g., to run light projectiles on heavy targets (as is now being effectively employed), in order to see the onset of phenomena with increasing nuclear size, was recognized from the beginning.    The 
initial luminosity was set at $10^{25}$ cm$^{-2}$sec$^{-1}$ with a possible upgrade 
to $10^{28}$ cm$^{-2}$sec$^{-1}$.   With RHIC inheriting six intersection regions from the CBA, the Task Force decided on having two major experiments, now PHENIX and STAR, and at least one small one, which were BRAHMS and PHOBOS eventually.  A frantic summer culminated in the third Quark Matter meeting at BNL in Sept. 1983, which galvanized the ultrarelativistic heavy ion physics community into planning for RHIC.

   The formal RHIC proposal of 1984 optimistically foresaw first beams in 1990, but approval of the funding agencies was slow and first funds for RHIC construction would be delayed until 1990.    From 1985 to 1995 the overall direction of the program was overseen by the RHIC Policy Committee.  Given the need to use the limited funds available to ensure that detectors would in place when the collider construction finished, little attention was paid initially to having a theory program in conjunction with the collider.  This gap would be well filled by the RIKEN Brookhaven Research Center, conceived in 1995 and begun in 1997, with T. D. Lee as the first director.  The first collisions at RHIC, Au on Au, were in 2000, a long 17 years after the Aurora meeting.
   
     Thoughts on colliding heavy ions  at CERN were reported by G. Cocconi at Bear Mountain \cite{bearmountain} who imagined transferring heavy ions, $^{16}$O to possibly U, from the PS to the ISR, and possibly into the SPS, then being built.  While the ISR, which ran from 1971 to 1984, could have been the first heavy ion collider (at 15 GeV/A) it was decommissioned in favor of LEP.  \, CERN actually began its fixed target program at the SPS in 1986, and reported first results, along with the AGS, at the Quark Matter 6 meeting in Nordkirchen in 1987.   The foundations of the LHC, itself, were laid down at the workshop of the European Committee for Future Accelerators (ECFA) at Geneva and Lausanne in March 1984, and the possibility of accelerating heavy ions considered early on, by H.  Specht and J. Sch\"ukraft among others \cite{rafelski}.   At the Evian-les-Bains meeting in March 1992  Sch\"ukraft presented an Expression of Interest for a dedicated experiment, and following a Technical Proposal in 1996, the  ALICE detector was approved in 1997 \cite{schukraft}.   The first experimental collisions of Pb on Pb at the LHC, at 2.76 TeV/A, took place in Nov. 2010, an even longer 26 years after the ECFA-CERN meeting. 
     
     \newpage

\section{QCD phase diagram of dense hot matter}

\begin{wrapfigure}[12]{r}{4cm}
\includegraphics[width=4.0cm]{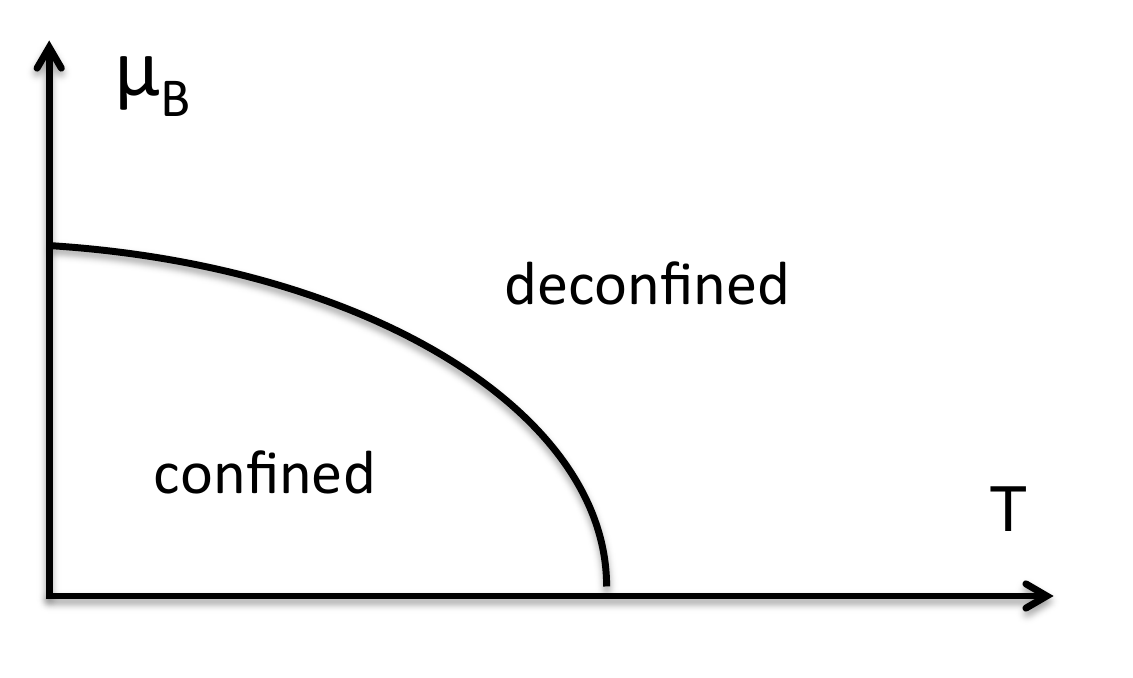}
\caption{
\footnotesize{
 Cabbibo-Parisi phase diagram of hot dense matter.}
}
\label{cabbibo}
\end{wrapfigure}

\begin{figure}[h]
\vspace{0cm}
\begin{minipage}{0.5\hsize}
\vspace{-0.18cm}
\hspace{.0cm}
\includegraphics[width = 1.0\textwidth]{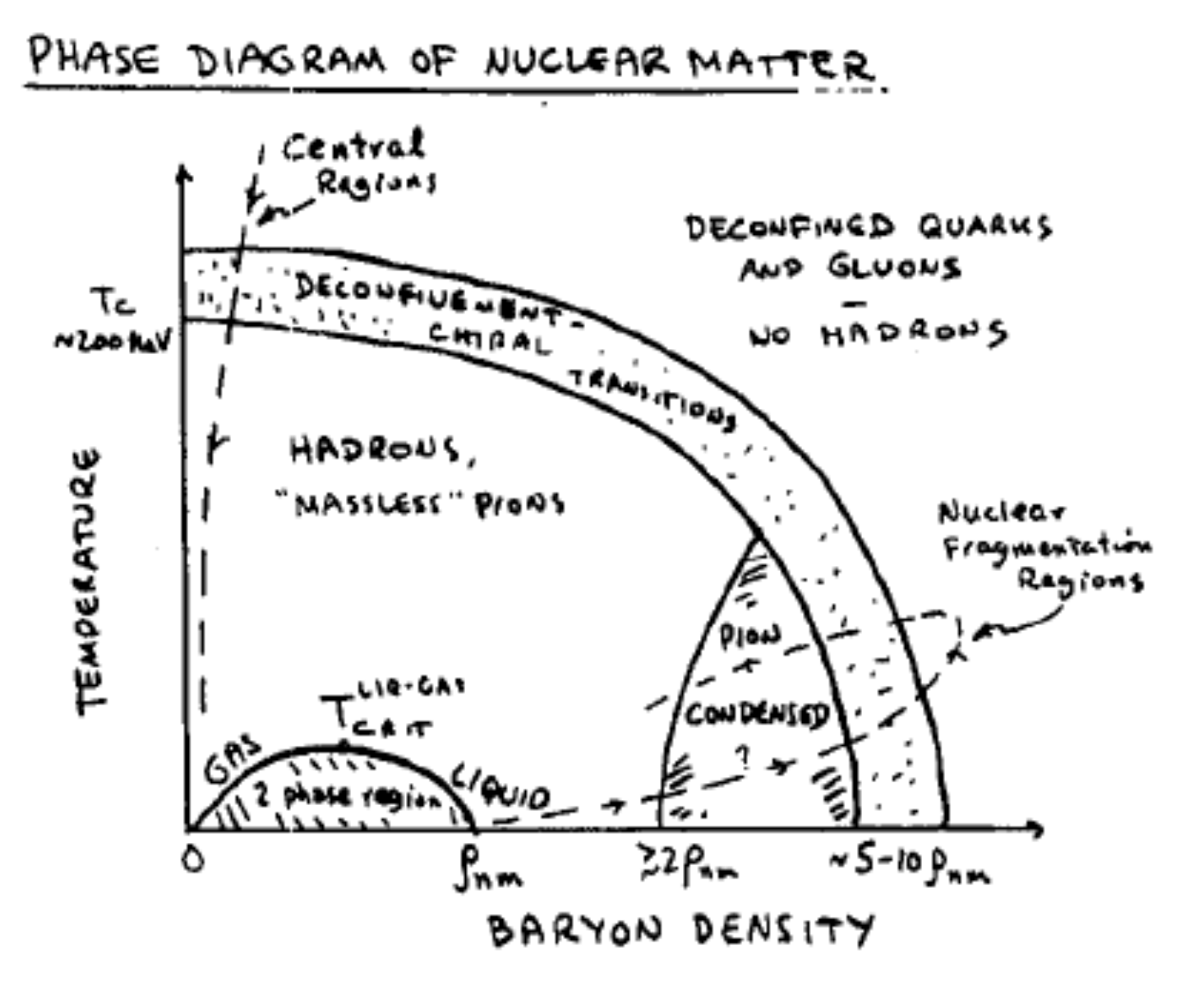}
\end{minipage}
\begin{minipage}{0.5\hsize}
\vspace{1cm}
\hspace{0cm}
\includegraphics[width = 1.15\textwidth]{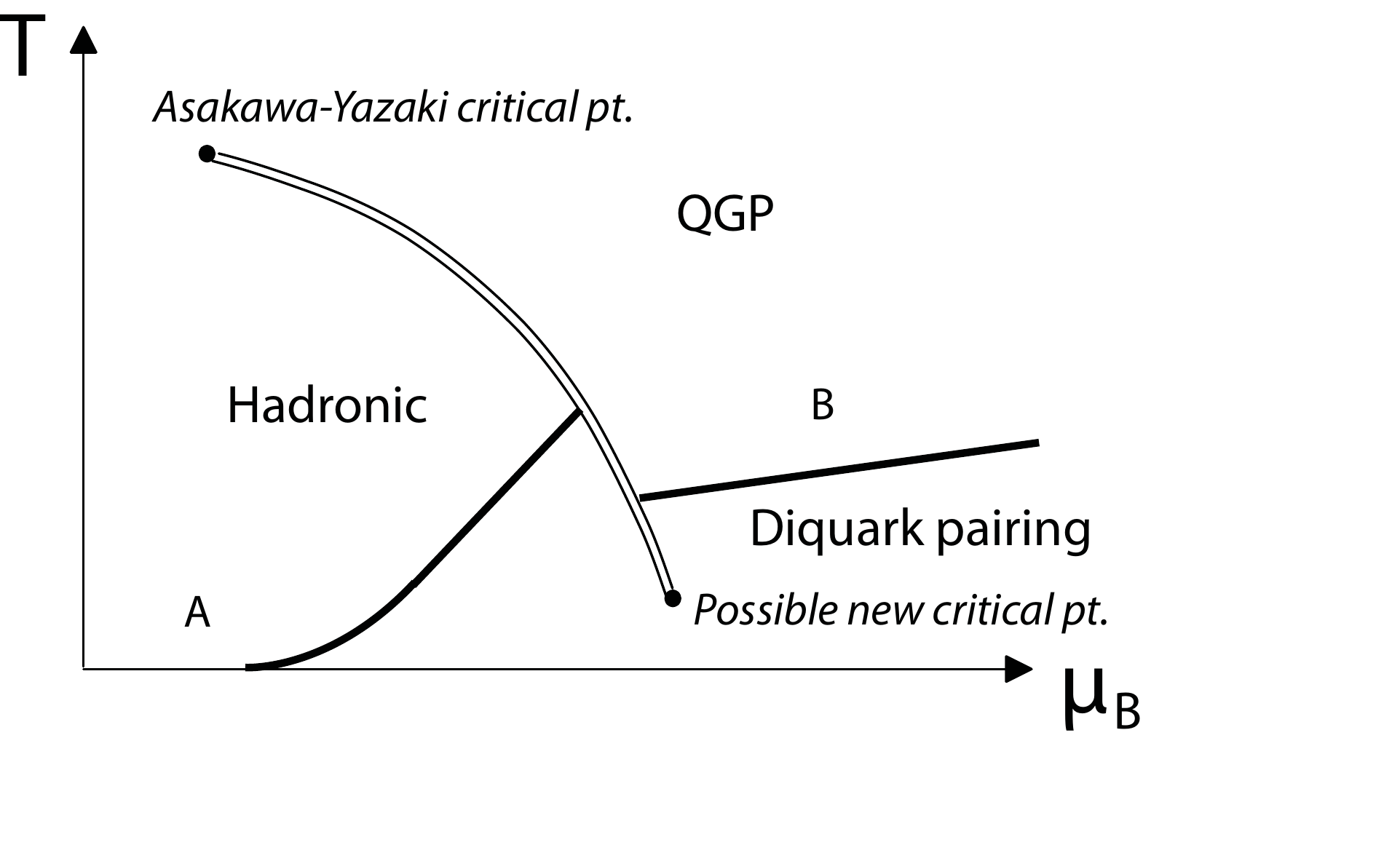}
\end{minipage}
\vspace{0cm}
\caption{
\footnotesize{
 Left:  Original phase diagram in the 1983 NSAC Long Range Plan.  \quad Right: Schematic modern phase diagram, showing the upper Asakawa-Yazaki
 critical point, a possible new critical point, and the region of diquark pairing.  The double line is a first order transition, while the solid lines
 delimit the region of coherent diquark pairing, or condensation.  Low temperature states of BCS pairing of neutrons or protons is not shown.
 }
}
\label{phases}
\vspace{0.2cm}
\end{figure}

   Let me turn now to physics developments.  The first construction of a phase diagram for matter at high density and temperatures, redrawn in Fig.~\ref{cabbibo}, dates back to Cabbibo and Parisi \cite{cabbibo} who simply
divided the temperature $T$--baryon chemical potential $\mu_B$ plane into two regions, confined at low temperature and density, and
deconfined elsewhere -- with a second order phase transition between the two.    Below, in the left panel of Fig.~\ref{phases}, is the phase diagram
I sketched for the 1983 NSAC Long Range Plan \cite{nsaclrp} showing regions in the $T$- baryon density plane explored by heavy ion collisions and neutron stars, as well
as the region of expected nuclear pion condensation.
The thick band between the hadronic and quark-gluon plasma regions was
intended to show uncertainties in the transition.   In the intervening years, many refinements have been made to the phase diagram.  These  include the high temperature, low chemical potential Asakawa-Yazaki critical point \cite{askawa}, whose detection is the subject of ongoing beam energy scans at RHIC and the
SPS. (discussed in detail elsewhere in this volume); 
possible states of diquark or Bardeen-Cooper-Schrieffer (BCS) pairing; and a possible low temperature critical point arising as
a consequence of the QCD axial anomaly breaking axial U(1) symmetry \cite{yamamoto}.  These features are sketched in the right panel of Fig.~\ref{phases}.  A more detailed version of this phase diagram is given in Ref.~\cite{fh}.

    At low temperatures  and high densities quark matter is expected to be a BCS color
superconductor, as a consequence of attractive gluon exchange in the  $^1$S$_0$ color and flavor anti-triplet channel
  \cite{colorsup,alford}.  The most energetically favorable diquark or BCS pairing states are the two-flavor 2SC or
isoscalar state, in which only $u$ and $d$ quarks are paired, and the color-flavor locked (CFL) state, which is the most stable for three equally populated flavors
of massless quarks in the weak coupling limit, both at $T=0$ and
near the critical temperature for the onset of pairing \cite{iida}.  This latter phase, which
breaks both color and flavor symmetry, contains 18 condensates, e.g., pairing of a red $u$
quark with a blue $s$ quark, a blue $s$ quark with a green $d$ quark, etc.    Diquark pairing ceases at sufficiently large temperature; in addition at lower $\mu_B$ diquark condensation ends as the strength of correlations among three quarks that eventually become baryons grows, since when a diquark carries around an extra unpaired quark the Pauli principle prevents the diquark from having a macroscopic probability of having zero momentum.  
This transition is denoted by the solid line to the right of point A in the right panel of Fig.~\ref{phases}.  

  While diquark pairing is unlikely to occur at high enough temperatures to play a role in ultrarelativistic heavy ion collisions,  
such pairing is significant in the structure of neutron stars, softening the equation of state at lower densities, and in the dynamics of neutron stars.  The 2SC and CFL phases respond quite differently to magnetic fields and rotation.  In an ordinary magnetic field, the 2SC state can form magnetic vortices 
with flux quantum $6\pi\hbar c/\sqrt{3g_s^2+e^2}$, where $g_s$ is the QCD coupling
constant and $e$ the electron charge.   As in a
rotating superconductor, this state responds to rotation by forming a very weak London
magnetic field which is a superposition of a weak electromagnetic magnetic field, $B \lesssim 1$ G, 
and a dominant color-magnetic field.   On the other hand, the CFL phase forms U(1) vortices in response to
rotation, as do superfluid neutrons, with a quantum of circulation,
$3\pi\hbar c^2/\mu_B$ three times larges than for neutron vortices \cite{iida3}.  Given this difference in the quanta of circulation, a join between vortices in the paired neutron phase
and those in the deconfined phase has to take place with three neutron vortices merging into a single quark vortex via a boojum \cite{boojum}.  Paired neutrons do not join continuously onto CFL paired quarks.

     The QCD axial anomaly leads to
an effective six quark Kobayashi-Miskawa-'t Hooft interaction, $\sim {\rm det}_{ij}\langle \bar
q^j_R q^i_L\rangle$, where $i$ and $j$ are flavor indices, and $R$ and $L$ are quark chiralities.
This interaction in turn attractively couples the chiral condensate and diquark pairing
fields of quarks in dense matter; chiral condensation favors diquark pairing, and vice versa.  It also opens the possibility, in matter with three flavors of quarks, of a low temperature critical point 
terminating the line of first order phase transitions, as shown in Fig.~\ref{phases} (right) \cite{yamamoto}.
Such a critical point is consistent with quark-hadron continuity, that matter crosses over smoothly from the
hadronic to the quark phase \cite{schaefer}, discussed below.
Whether or not such a critical point actually exists in the phase diagram
depends in detail on the strength of the six quark interaction in the strongly coupled regime \cite{pnjl}.   Were it to exist it would be at too cold a temperature to be 
experimentally accessible at RHIC or the LHC, but could possibly be reached at FAIR.

     Other possibly important features in the phase diagram, in the region of the low temperature critical point in Fig.~\ref{phases} (right), are spatially ordered chiral phases \cite{fh}, a form of {\em quark ice}, and a quarkyonic phase \cite{quarkyonic,sasaki}.

    The emerging role of quark-hadron continuity in the phase diagram can be seen in the crossover between the hadronic phase and the quark-gluon plasma for matter with low baryon chemical potential $\mu_B = 0$ at finite temperature $T$. 
As C. Ratti discusses more fully in her talk in this volume \cite{ratti} recent lattice gauge calculations indicate that
matter there undergoes a rapid crossover from a hadronic to a quark-gluon phase at temperature $T\sim$150-155 MeV with gradual and continuous restoration of chiral symmetry \cite{WB,hotQCD}.   (Were the light quarks massless, then at very high temperatures chiral symmetry would not be broken, and thus one would expect a second order chiral symmetry restoring phase transition instead of the crossover found for realistic quark masses.)

    Naively one might say that since quarks are free in the plasma phase above the crossover, by continuity free quarks would have a probability to be present in matter below the crossover, e.g., there could be free quarks running around in air, similar to the way there is a very tiny probability of free electrons being in air, from thermal ionization.  But since there cannot be free quarks in low density matter, the correct conclusion, as pointed out by T. Kojo, is that even above the crossover, there are no free quarks (except in the very high $T$ asymptotically free regime); rather the matter must consist of complicated clusters of gluons and quarks both above and below the crossover; such clusters are illustrated in his talk in this volume \cite{toru-nstars}.  The crossover, and deconfinement more generally, is characterized by a {\em percolation} transition, as first proposed in \cite{perc} for dense matter at zero temperature  and further amplified by H. Satz and coworkers \cite{satz}; also \cite{lottini}.   At the lowest temperatures, with $\mu_B \ll m_n$ (the nucleon mass), the clusters are isolated as single thermal pions, which become more and more connected as $T$ increases, through the gluon and quark exchanges responsible for the interactions of the pions, until the clusters fill enough of space that a single quark can propagate from one end to the other.  
 
  The regions of space in which quarks can move around are always net
color singlets.  At the percolation transition the sizes of the color singlet regions change from always being finite at low temperatures to being the size of the entire system, e.g., the collision volume in a heavy-ion collision.   Since one can go continuously around the first order transition line
between the hadronic and plasma phases, e.g., from point A to B in the schematic phase diagram in Fig.~\ref{phases} (right), these two phases are continuously related.  The first order transition line between hadronic matter and the quark-gluon plasma is closer to that in a liquid-gas phase transition, where one can go continuously from liquid to gas around the critical point  (in water at 373 C).    At lower temperatures at high density the matter undergoes nucleonic pairing and then quark pairing, phases distinct from that at higher temperature phase without pairing.   Scenarios of the evolution of cold matter is discussed in more detail in Refs.  \cite{sasaki,kenjitoru}.

     One strong initial motivation of the ultrarelativistic heavy-ion physics program was to learn about matter in neutron stars.  Although one cannot directly access matter at sufficiently low temperatures in collisions to learn directly about matter in neutron stars, it is quite remarkable how the study of matter at small $\mu_B$ and finite temperature, which is the regime in which one can begin to compare experiment with lattice gauge theory, is in fact informing us about the states of matter at large $\mu_B$ and low temperatures.  I return to cold matter below.

\section{Connections of ultrarelativistic heavy ions with other areas of physics}

    As envisioned in the early conception of the ultrarelativistic heavy ion program, the physics of collisions has had fertile connections with other fields of physics, including ultracold atoms (a field only imagined a billion seconds ago), cosmology and astrophysics.   
    
\subsection{Ultracold atomic physics}

    While at opposite limits of matter under extreme conditions of energy, ultracold atomic systems at temperatures of order microkelvin to nanokelvin ($10^{-13}$ eV) have many similarities with ultrarelativistic heavy ion collisions, despite there being some twenty orders of magnitude difference in their energy scales.  Both systems are small clouds with some $10^4$ or more degrees of freedom.  While quark matter in collisions is always strongly interacting, through Feshbach resonances one can adjust external magnetic fields to make clouds of cold atoms also strongly interacting.  Both systems become essentially scale free: in atoms in the resonance, in the {\em unitarity} regime, the two-atom scattering length, $a$, is much larger than the interparticle spacing, and is thus an irrelevant parameter; the only scale is the interparticle spacing.   Similarly dense matter in equilibrium has essentially only the scales of temperature or chemical potential.   Cold atom systems can be bosonic, becoming Bose condensed at low $T$, or fermonic, becoming BCS paired superfluids at low $T$, or indeed mixtures of bosons and fermions.    As a consequence of cold atoms at unitarity being scale free, the ratio of the first viscosity, $\eta$, to the entropy density, $s$, is very low;  in atomic fermonic $^6$Li, $\eta/s \lesssim 0.4$  at unitarity \cite{cao},  comparable with that at RHIC $\lesssim 0.2$.  
    
\begin{figure}[h]
\vspace{0cm}
\begin{minipage}{0.5\hsize}
\vspace{0cm}
\hspace{.0cm}
\includegraphics[width = 1.0\textwidth]{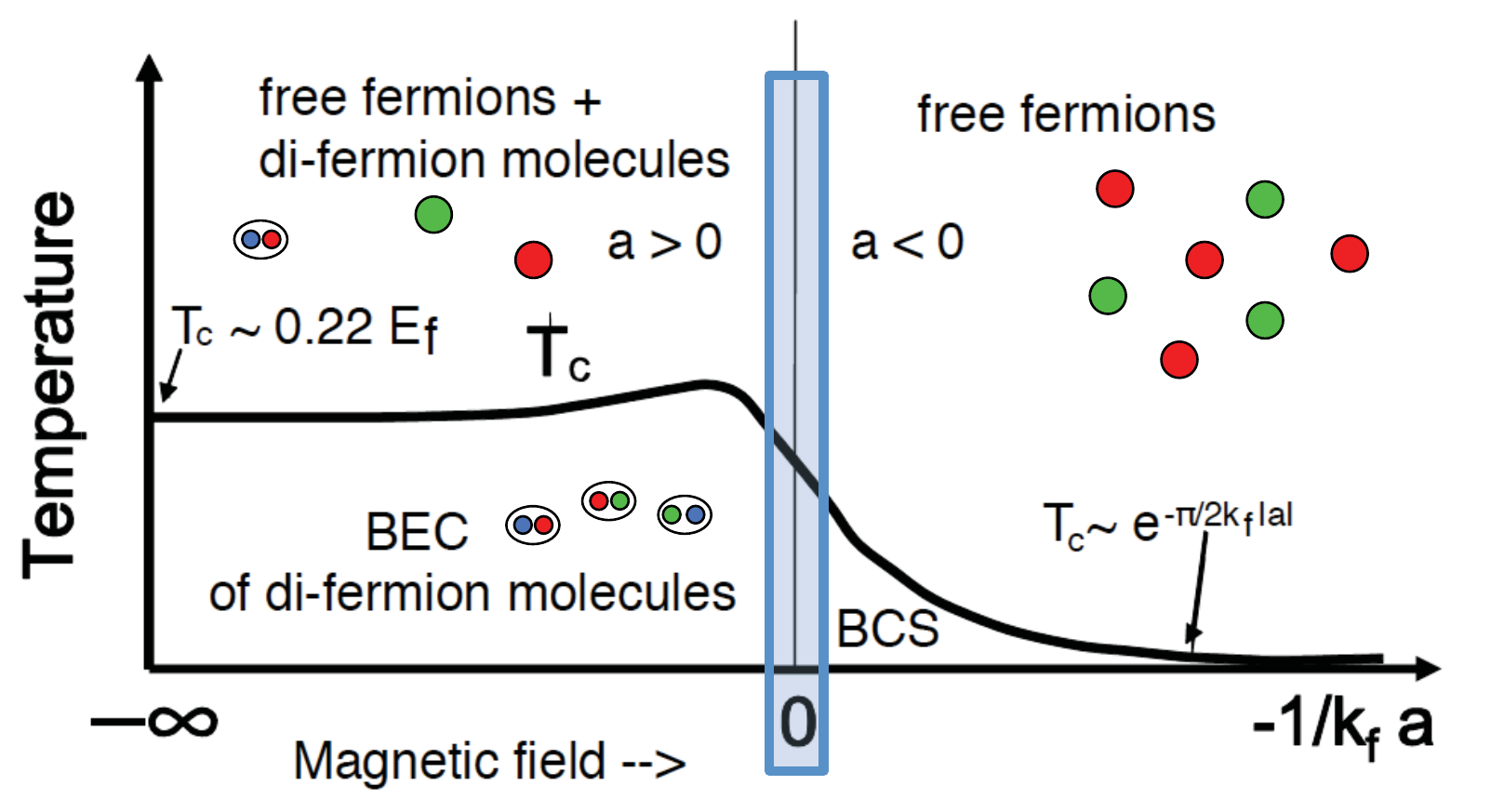}
\end{minipage}
\begin{minipage}{0.5\hsize}
\vspace{0cm}
\hspace{-.1cm}
\includegraphics[width = 1\textwidth]{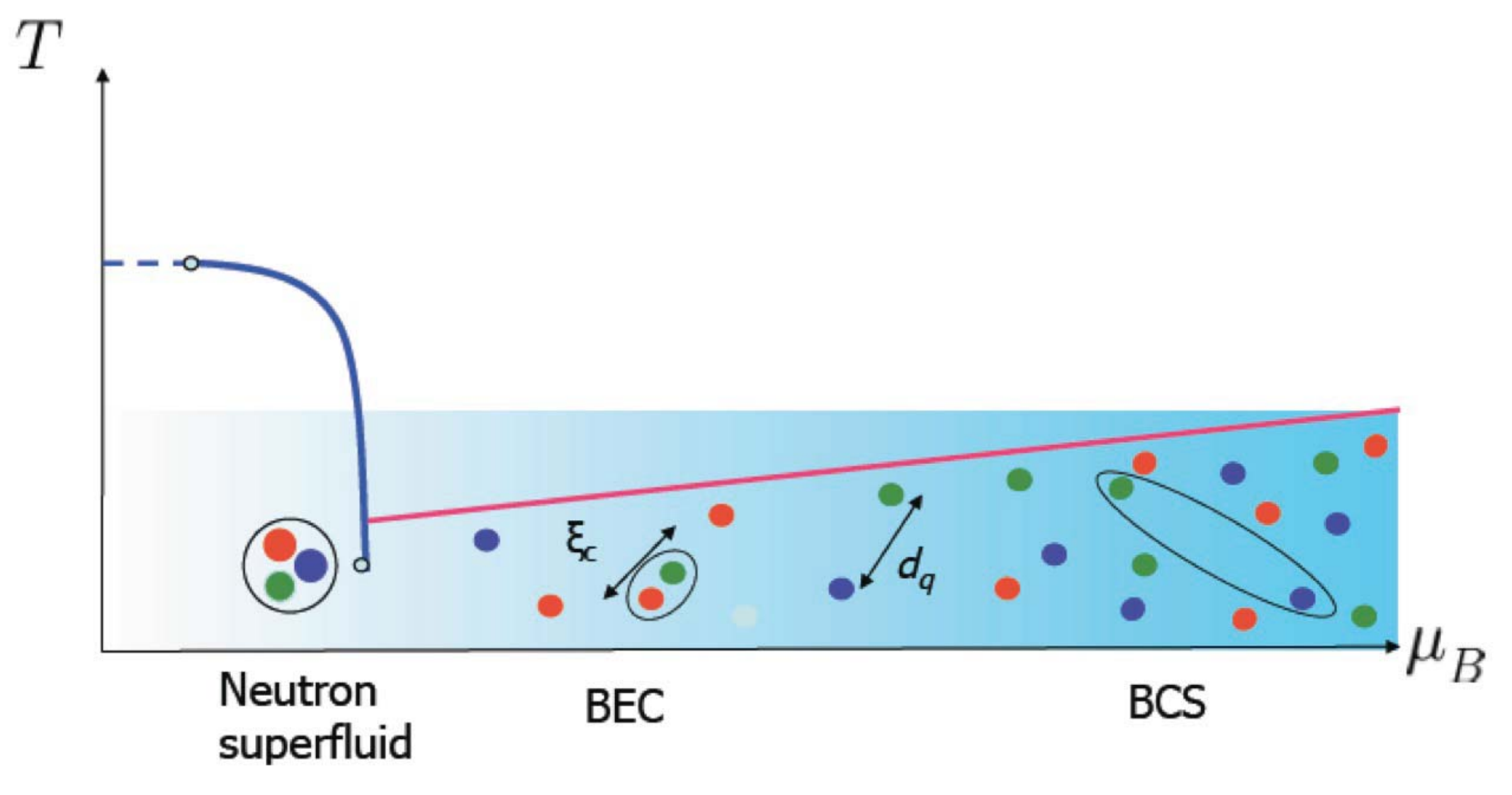}
\end{minipage}
\vspace{0cm}
\caption{
\footnotesize{
 Left:    Phase diagram of a gas of two equally populated hyperfine states 
of ultracold atomic fermions as a
function of the negative of the inverse scattering length $a$, in units of the
Fermi momentum.   The transition between the BEC and
BCS regions is a smooth crossover. The continuous curve is the finite temperature condensation transition, approaching the weakly interacting BEC transition
temperature to the left and the BCS transition to the right.\quad
 Right: Corresponding transition from a BEC of diquarks to BCS pairs in low temperature quark matter (after T. Hatsuda)..}
}
\label{bec-bcs}
\vspace{0.2cm}
\end{figure}
    
      Figure~\ref{bec-bcs} shows the phase diagram of fermonic atoms in two spin (or hyperfine) states as a function of $-1/k_Fa$, where the Fermi momentum $k_F$ is given by $(3\pi n_{atoms})^{1/3}$.    To the left the atoms at high temperature are free, and with decreasing $T$ they form molecules which eventually Bose condense.  To the right the atoms at high temperature are also free, but
with decreasing $T$ they undergo a BCS phase transition.   A remarkable feature is that the crossover at low temperatures -- as the scattering length goes through infinity, from a Bose-Einstein condensate (BEC) of diatomic molecules for weak positive $a$ on the left, 
to a BCS paired superfluid on the right -- is completely smooth.   The molecules continuously expand
in size from tightly bound in the BEC regime to widely spaced pairs in the BCS regime.  This crossover suggests similar behavior in dense matter,
that the quarks go continuously from being BCS paired at very large $\mu_B$ and low $T$, to a BEC of diquarks at lower $\mu_B$, as shown in the right panel of  Fig.~\ref{bec-bcs}.   Were the color group just SU(2),  the analogy would be more complete; the diquark pairs would be the hadrons.   Understanding how matter with SU(3)-color crosses over
from having strong three-quark correlations at low densities to two-quark correlations at high densities remains an open issue.   
Since at lower $\mu_B$, two-flavor 2SC color pairing is favored,
the atomic systems suggest that at low $T$ three-quark
correlations could enter through BCS pairs shrinking into diquarks with decreasing density, turning continuously into a strongly
interacting diquark BEC;  the condensed pairs  should eventually bind
to the unpaired quarks to form baryons at lower density, with a loss of their Bose-Einstein condensation en route.   One can simulate this latter transition in a mixture of ultracold bosons (the analogs of diquarks) and fermionic atoms (analogs of the unpaired quarks) binding into molecules, the analog of the nucleon~\cite{kenji}.  More generally, atomic fermionic systems with three
internal states, e.g., the three lowest hyperfine levels of atomic $^6$Li, could enable laboratory analogs of the QCD 
deconfinement transition with pairing correlations.\cite{3levels}.

    The growing ability to simulate magnetic fields in cold atom systems opens an important connection of cold atoms to dense matter.  One is learning how, through very clever combinations of external lasers \cite{simulated}, to make neutral atoms behave similarly to electrons in magnetic fields, and go around in circles (or vortices).  Furthermore, one can begin to simulate non-Abelian gauge fields, so far limited to producing artificially spin-orbit coupled atoms \cite{spinorbit}.   At the moment, the artificial fields can only be made to be slowly varying over the system; one is not yet able to construct short wavelength addressable fields.  Furthermore, while the atoms respond to the external fields one is not yet at the point where the atoms themselves can modify the fields, i.e., one does not have polarization effects.   Eventually, however, one can look forward to the very exciting possibility of carrying out analog simulations of QCD lattice gauge theory using  cold atom systems with many internal (hyperfine) degrees of freedom \cite{wiese}. 

     A few other connections, old and new:  As one has learned from cold atom experiment with two unbalanced hyperfine states \cite{mz}, pairing tends to push the excess population out of the region where pairing takes place, shedding possible light on how, in degenerate cold quark matter with fewer strange than light quarks,  e.g., in neutron stars, $s$ quarks can BCS-pair with $u$ or $d$ quarks.     A very fertile approach which has great promise for simulating strongly interacting plasmas including their dynamical evolution and instabilities is to use ultracold table-top ionized atomic plasmas \cite{atomplasmas}.   Finally, if gluons in the early states of ultrarelativistic collisions become overpopulated as a consequence of the slowness of gluon-annihilating collisions, they can form a Bose-Einstein condensate \cite{blaizot}.  Understanding the dynamical evolution and indeed possible signatures of such a condensate is every bit as challenging as in cold atom physics.
    
\subsection{Cosmology and astrophysics}

    The finite temperature, low $\mu_B$ crossover predicted by lattice QCD  has the immediate implication for cosmology that the transition from the quark-gluon plasma prior to the first microsecond of the early universe to later hadronic matter was smooth.  Thus one should not expect to see signatures of a first order hadronization transition, such as effects of deflagration bubbles on the large scale structure of the universe, and the 
formation of planetary size black holes  (reviewed in \cite{earlyuniv}).   In the other direction, the recent interest in the possibility of observing the imprint of primordial gravitational radiation and other anisotropies in the early universe on the polarization of the cosmic microwave background \cite{adam} has raised the question of using polarization of real and virtual photons (dilepton pairs) produced in ultrarelativistic heavy ion collisions to probe the anisotropies in the early stages of collisions \cite{polarizedHIC}.   Indeed, the anisotropy introduced by the collision axis makes the collision volume act effectively as a birefringent emitter, with the extraordinary axis along the beam direction.  Detecting the polarization of direct photons remains a difficult experimental task, while detecting that of  dilepton pairs is more feasible \cite{specht}.

\subsection{Neutron stars}

   \begin{figure}[h]
\vspace{0cm}
\vspace{6pt}
\hspace{15pt}
\includegraphics[width = 0.4\textwidth]{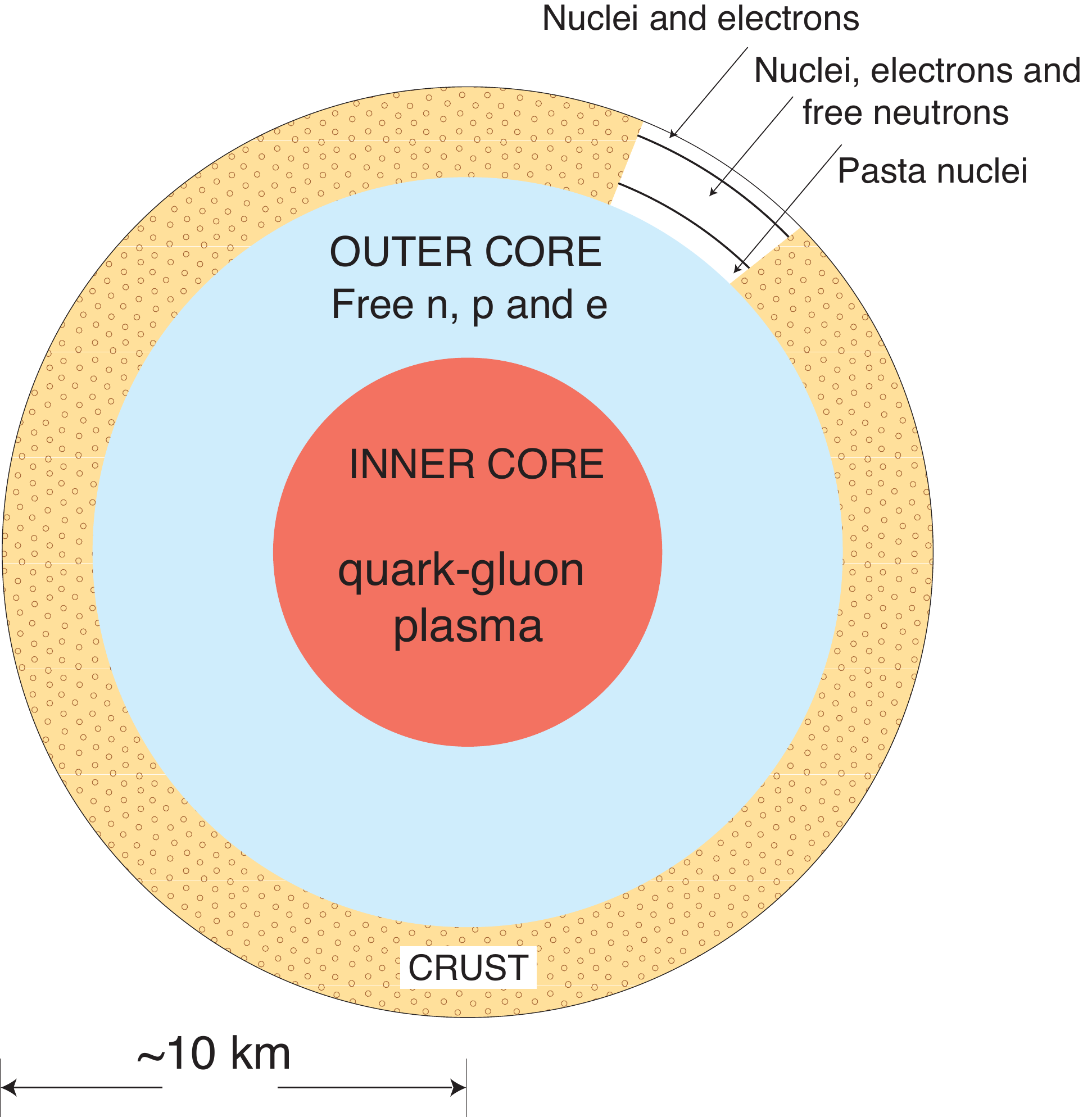}
\vspace{0cm}
\caption{
\footnotesize{{
Cross section of a neutron star showing the crust, nuclear matter in the outer core, and the quark matter inner core.}
}
}
\label{nstar}
\vspace{0.2cm}
\end{figure}

    As mentioned, learning about matter in the deep interiors of neutron stars  (Fig.~\ref{nstar}) was a prime motivation for carrying out heavy ion collisions, and indeed neutron stars provide a complementary astrophysical laboratory for determining the properties of dense matter.   Unfortunately, collisions cannot (so far) probe matter at the low temperatures, $\lesssim$ 1 MeV, and high baryon densities, $n_B$, in neutron stars, some 5-10 times normal nuclear matter density, $n_0$; the present inability of lattice gauge theory to calculate directly and reliably at finite $\mu_B$ is a corresponding theoretical obstacle.  Thus deducing neutron star properties requires constructing a picture of their interiors fitting together neutron star observations with theoretical pictures of cold dense matter that are consistent with what we are learning about dense matter at high temperatures, importantly quark-hadron continuity in the phase diagram.
   
   \begin{figure}[h]
\vspace{0cm}
\begin{minipage}{0.5\hsize}
\vspace{6pt}
\hspace{15pt}
\includegraphics[width = 0.8\textwidth]{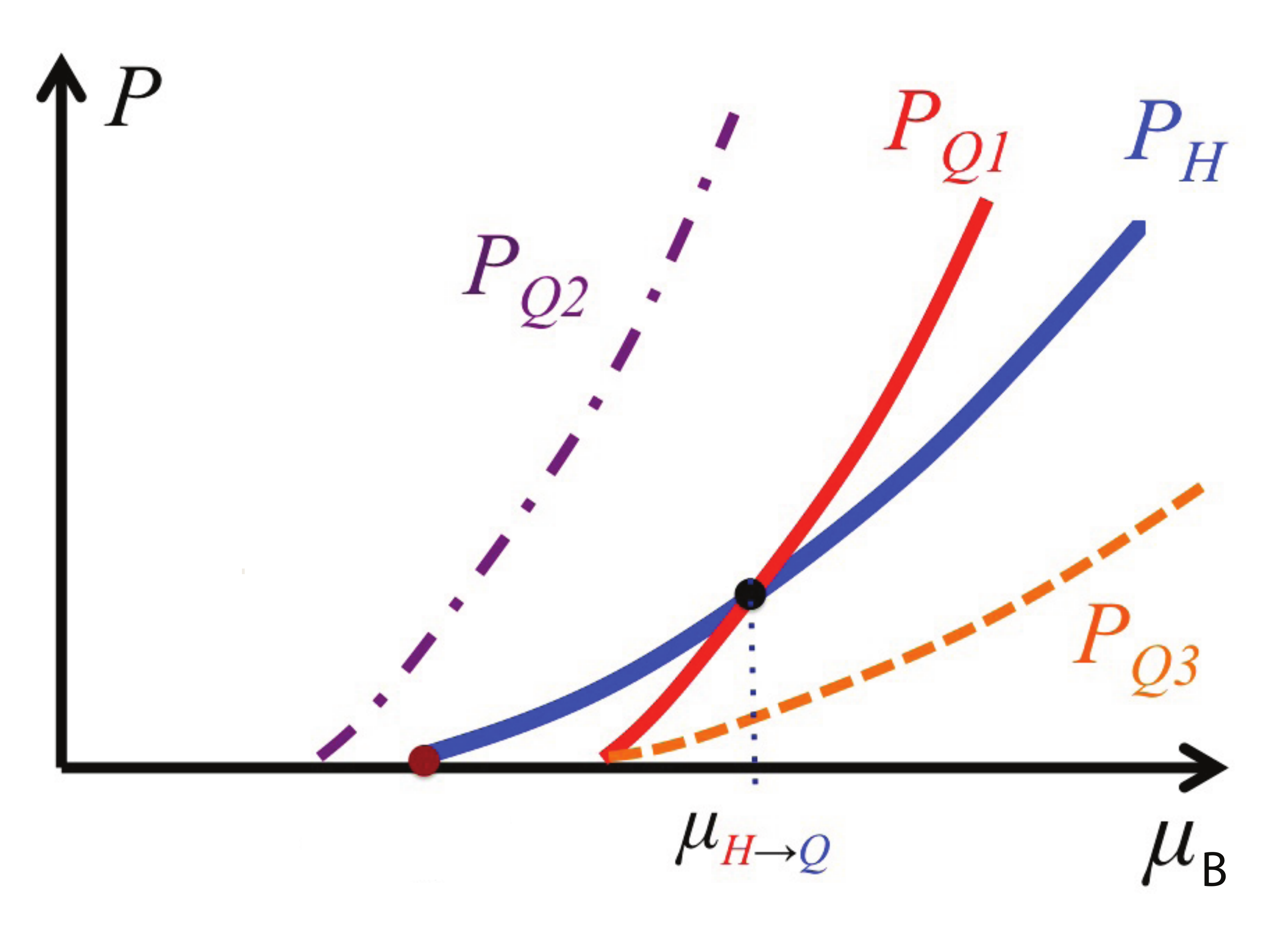}
\end{minipage}
\begin{minipage}{0.5\hsize}
\vspace{2pt}
\hspace{18pt}
\includegraphics[width = 0.8\textwidth]{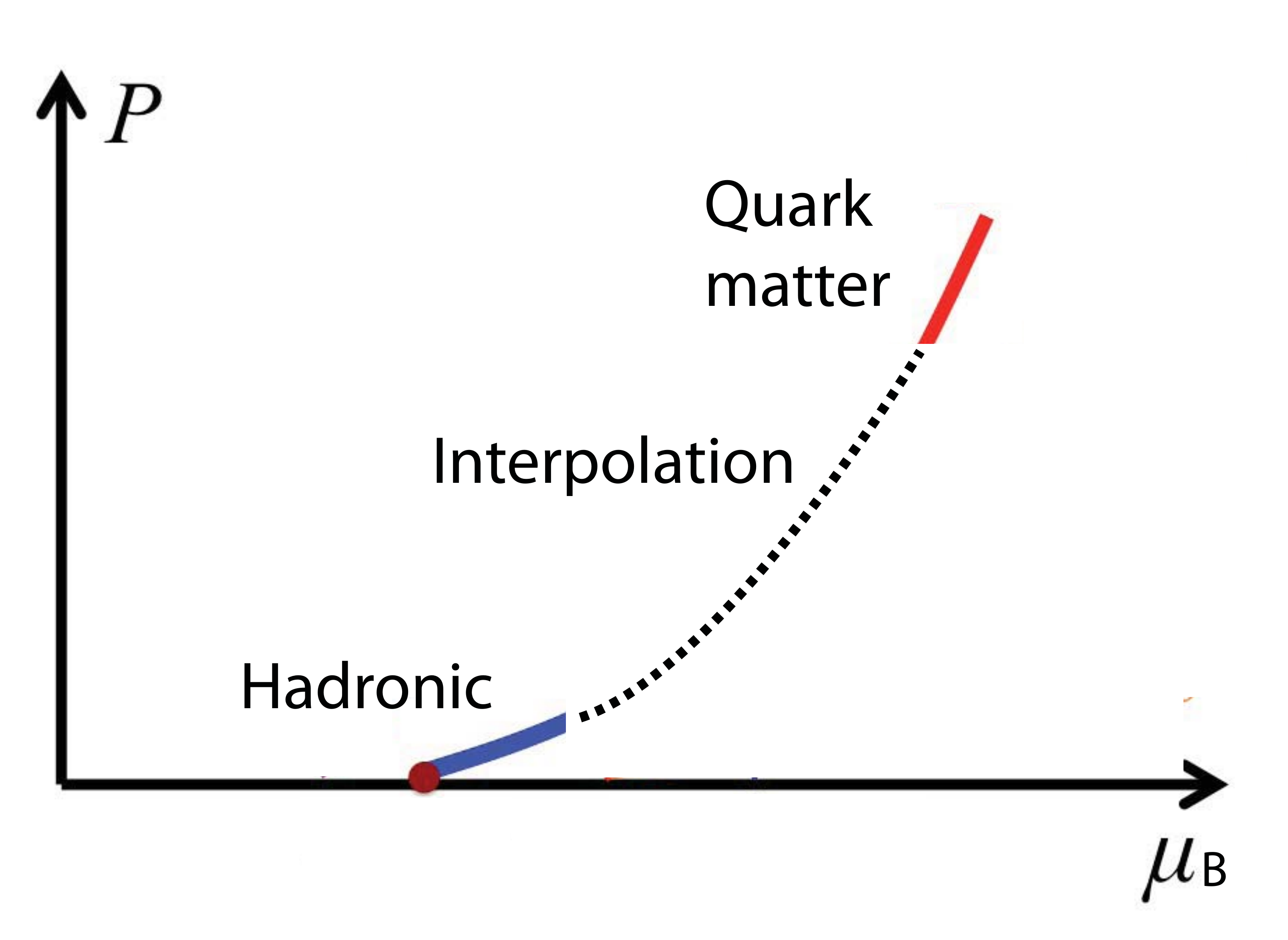}
\end{minipage}
\vspace{0cm}
\caption{
\footnotesize{
 Left: Pressure vs. baryon chemical potential in the usual construction, showing the crossing of the quark pressure $P_{Q1}$ with the hadronic pressure $P_H$, and a phase transition at $\mu_{H\to Q}$.   Stiff quark equations of state such as $P_{Q2}$ and $P_{Q3}$ that do not cross the hadronic curve are excluded.   \quad
 Right:  Smooth interpolation of the pressure between hadronic matter at low $\mu_B$ and quark matter at high $\mu_B$, which
accommodates much stiffer quark matter equations of state. }
}
\label{pmu}
\vspace{0.2cm}
\end{figure}

    A fundamental problem is whether quark matter exists in neutron star cores. The standard description of the onset of quark matter, used first in \cite{chin}, assumes that matter at high density can be in either a hadronic or quark phase; the favorable phase has the lower energy density at given $n_B$, or equivalently has the  higher pressure, $P$, at given $\mu_B$.  To illustrate, in Fig.~\ref{pmu} (left) the hadronic state $P_H$ in this construction wins at lower $\mu_B$ and the quark state $P_{Q1}$ is favored at higher $\mu_B$. with a first order transition at $\mu_{H\to Q}$.   However,  this method has a serious conceptual problem in that it assumes the existence of nucleonic matter ($P_H$) at too high a density to be physical, as well as comparing nucleonic matter there with quark matter ($P_{Q1}$) at a density too low  to apply perturbative QCD.  The procedure thus excludes stiff quark equations of state such $P_{Q2}$ and $P_{Q3}$ that do not cross the hadronic pressure curve.   As a consequence, neutron stars constructed from such an approach generally have at most a small quark core inside a hadronic liquid, see, e.g., \cite{APR}.   
   
    The recent observation of two massive neutron stars, PSR J1614-2230 with $M =1.97 \pm 0.04$ solar masses ($M_\odot$) \cite{Demorest} and PSR J0348+0432 with $M =2.01 \pm 0.04 M_\odot$ \cite{Antoniadis2013}, both in binary orbits with stable white dwarfs, tells us immediately that the equation of state of dense cold matter must be very stiff, and immediately raises the question of whether quark matter can be stiff enough to support such massive stars.    The answer is yes!   The requirements that the equation of state describes nuclear matter at low densities, $\lesssim 2-3\, n_0$, and quark matter at high, $\gtrsim 7-8\, n_0$,  and that the quark matter be sufficiently stiff to allow neutron stars of at least 2$M_\odot$ strongly constrains the possible intervening equation of state.  The most physical approach then is to consistently interpolate the pressure as a function of $\mu_b$ between these limits, as shown in Fig.~\ref{pmu} (right) \cite{toru-nstars,toru-nstars1,masuda}, an approach compatible with quark-hadron continuity, and also with a possible first order phase transition.   As described in detail in T. Kojo's talk in this volume \cite{toru-nstars}, one can with reasonable strengths of a vector repulsion between quarks and the diquark pairing interaction readily construct high mass neutron stars with extensive 
quark cores.   The resulting equation of state is consistent with that deduced phenomenologically from simultaneous determinations of masses and radii, $\sim 10-12$ km, for some dozen neutron stars \cite{feryal}.  Although numerically similar to the nucleonic-based APR equation of state \cite{APR}, such an interpolated equation of state is not based on the unphysical assumption that nucleons are tremendously jammed together in the deep interior of neutron stars, yet continue to interact via two- and three-body forces as they do in ordinary nuclei.

   Quark matter equations of state can be stiff enough to support neutron stars of masses $\gtrsim 2 M_\odot$.
In the regime, $2n_0 \lesssim n_B \lesssim 7-8 n_0$, matter is intermediate between purely hadronic and purely quark, perhaps quarkyonic \cite{kenjitoru}.  Despite uncertainties remaining in this picture -- from the microscopic descriptions of the nuclear matter in beta equilibrium at lower densities to the intermediate regime to the quark matter at high densities -- we are now evolving a picture of dense matter in neutron stars consistent with theoretical expectations of its properties deduced from ongoing lattice calculations and ultrarelativistic heavy ion collisions.

\section{Conclusion}

    To what extent have we reached our original goals?   The ultrarelativistic heavy ion program has succeeded in producing nuclear matter at energy densities several orders of magnitude beyond that in nuclei at rest.  It has certainly discovered a new state of  matter, the quark-gluon plasma, which is strongly interacting, and behaves collectively.  Although it was well realized that the plasma in heavy ion collisions would be strongly interacting, its manifestation in terms of its very small viscosity to entropy ratio, consistent with the AdS/CFT bound, was came as a lovely surprise, as nature is indeed wont to present.   The intellectual challenges of ultrarelativistic heavy ion physics continue to inspire remarkable theory, from the early stages of collisions to the onset and evolution of hydrodynamics to its thermodynamics, its quasiparticle structure and collective modes \cite{blaizot-qp}. 
 Our knowledge of the phase diagram is undergoing continual refinement, with beam energy scans searching for the Asakawa-Yazaki critical point.   The lattice calculations at $\mu_B$ = 0, in indicating a crossover, are giving new insights into the continuity of matter in the phase diagram, and point to the validity of the percolation picture.  The parallel development of understanding how quark matter can support massive neutron stars, as observed, further informs this picture.      Connections of ultrarelativistic heavy ion collisions to other fields continue to be made, in both expected and unexpected ways, e.g., the development of the chiral magnetic effect in collisions \cite{dima} and its application to condensed matter systems \cite{li}, work that begins to relate the role of topology in condensed matter to heavy-ion physics; and the future possibilities of simulations of non-Abelian fields in QCD with cold atoms.   As this conference makes clear,  experimentalists -- with considerable theoretical backing -- have succeeded very far beyond our first expectations back in Aurora in bringing out the physics of heavy ion collisions.   The story fortunately is not over \cite{white}.

\section*{Acknowledgments}    I am very grateful to the organizers, H. Hamagaki, Y.  Akiba, and T. Hatsuda, as well as RIKEN, for giving me this chance to look back over the first billion seconds of ultrarelativistic heavy ion collisions.   I also thank J.-P. Blaizot, P. Braun-Munzinger, K. Fukushima, T. Hatsuda, T. Kojo, and J. Stachel for  very helpful discussions during the Kobe meeting about the nature of the phase diagram, and H. Bech Nielsen  and
W. Weise for subsequent discussions.   Work reported here was supported in part by the U.S. National Science Foundation, most recently by NSF Grant PHY13-05891.

\vspace{24pt}

\bibliographystyle{elsarticle-num}

\end{document}